# MODELING AND ESTIMATION OF THE RISK WHEN CHOOSING A PROVIDER

*E. Sorokina*
*Department of Mathematics, University of Technology, Korolev, Russia*
vorhtc256@gmail.com

**Abstract** - The paper provides an algorithm for the risk estimation when a company selects an outsourcing service provider for innovation product. Calculations are based on expert surveys conducted among customers and among providers of outsourcing. The surveys assessed the degree of materiality of species at risk.

The factors of the enterprise's economic stability are its competitive ability, quick response to the changes of external conditions, implementation of the new technologies, increase of the manufactured products quality and efficiency of the industrial enterprise activity. One of the methods of competitive ability increase for an enterprise is the use of outsourcing [1 - 4]. However, upon outsourcing implementation in the enterprise, it shall be accounted that snap decision about transition to outsourcing, not supported by a deep analysis of all potential risks [5 - 11] and advantages can have a negative effect on the enterprise activity. This is connected with the fact that the outsourcing market in our country is not yet developed to such a degree as to offer large suppliers to the enterprises, suppliers, who would be able to provide high-level services.

Transition of the enterprise's operation functions to outsourcing requires that corporate decision support systems (DSS) shall assess options of orders placement for outsourcing with account of risk factors [12, 13].

In this work, the algorithm is suggested for calculation of risk integral estimate upon selection of an outsourcing provider, a numerical example is provided.

Each company operating in one or another area of economics, usually has a current portfolio of outsourcing providers. The pool of providers can be stable or highly dynamic. Features of each of them can eventually vary. We can emphasize the following features:
- Provider's work experience (both positive and negative);
- Competitive advantages;
- Range of service prices;
- Provision of accompanying services;
- Flexibility in relations with the customer;
- Key performance indicators of a company providing outsourcing services;
- Personnel qualification;
- Feedbacks of the customers, etc. [10].

DSS shall have means of current assessment, including estimate of characteristics of risks [12 - 14] associated with each provider. In respect of the provider's risk characteristic, it is frequently recommended to have a generalized indicator integrating multiple factors (experience, reliability, flexibility, qualification, etc.).

Variety of risks accounted in different economy braches, is quite large [8 - 10].

In respect of the influence degree, the risks are divided into external (uncontrolled) and internal (controlled). External risks include:
- economic (price risk, exchange rate, currency and market risks);
- administrative (modification of the statutory documents, payments accompanying risks);
- risks associated with outsource services provider (breach of contract terms, information leak, growth of prices).

Internal risks are usually represented by information risk (untimely receipt of information), personal risk (associated with the low professional level of the decision-makers), financial risk (lack of funding).

Diversity of possible sources of information about risks complicates their comprehensive and complete accounting. Therefore, in practice, experts are frequently involved for the assessment of poorly formalized and hardly-measurable factors. In this work, expert evaluation tools are also applied for risk assessment of each provider. The suggested technology includes two groups of expert evaluations:

1. The first group includes estimates of relevance of each factor according to the customers' opinion – and separately according to the provider's opinion. The result is the balanced weights of each risk factor according to the opinion of such services market participants. This group can be formed by specialized consulting companies with the use of expert evaluation methods for different types of businesses used in the outsourcing (transport, legal, customs services, provision of constituent components and ingredients for different manufactures, etc.)

2. The second type is formed by the company selecting provider itself. Evaluations here are represented by the scores assigned by the experts for one or another risk factor, applicably to a certain provider from the portfolio of potential outsourcing services providers. For instance, even with a high risk of unreliable supply, a minimum score can be assigned, in case the company receives the commodities in form of customer's pick-up.

Algorithm of provider's risk evaluation is demonstrated by the model data provided in the Table 1.

**Table 1. Initial data and estimates**

| № | Risk factors | Estimates ("-score" ranges) | | | | | $c_i$ | $\alpha_i$ | $b_i$ | $\beta_i$ | $\gamma_i$ |
|---|---|---|---|---|---|---|---|---|---|---|---|
| | | 1 | 3 | 5 | 7.5 | 10 | | | | | |
| 1 | Experience | 0.01 | 0.02 | 0.07 | 0.07 | 0.18 | 9.22 | 0.18 | 1 | 0.07 | 0.10 |
| 2 | Image | 0.09 | 0.02 | 0.31 | 0.22 | 0.13 | 7.03 | 0.13 | 1 | 0.07 | 0.08 |
| 3 | The scale of production | 0.24 | 0.09 | 0.15 | 0.12 | 0.11 | 5.98 | 0.11 | 3 | 0.20 | 0.20 |
| 4 | The term of execution | 0.30 | 0.03 | 0.11 | 0.16 | 0.12 | 6.17 | 0.12 | 1 | 0.07 | 0.07 |
| 5 | Financial condition | 0.11 | 0.07 | 0.14 | 0.21 | 0.15 | 7.99 | 0.15 | 2 | 0.13 | 0.18 |
| 6 | The price of the service | 0.24 | 0.10 | 0.18 | 0.20 | 0.11 | 5.71 | 0.11 | 3 | 0.20 | 0.19 |
| 7 | The source of financing | 0.05 | 0.12 | 0.16 | 0.24 | 0.12 | 6.33 | 0.12 | 2 | 0.13 | 0.14 |
| 8 | National identity | 0.64 | 0.17 | 0.09 | 0.05 | 0.05 | 2.50 | 0.05 | 1 | 0.07 | 0.03 |
| 9 | Advertising activity | 0.93 | 0.02 | 0.03 | 0.01 | 0.03 | 1.35 | 0.03 | 1 | 0.07 | 0.02 |

Thick frame contains portions $q_{ij}$ of the respondents (interviewed customers and providers), whose 0-10 evaluations lie within the "-score" ranges [0; 1], [1; 3], [3; 5], [5; 7.5], [7.5; 10], right borders of which (pockets) are specified as $a_i$. These values can be calculated separately for the customers and providers, but here their average values are presented, i.e. the correlation ratio for these groups amounts to 0.94, which allows to consider their opinions about risk factors as consistent. Portions of respondents are obviously can be interpreted as probabilities of the corresponding scores $a_i$. Then it is possible to estimate average risk $c_i$ as an average score of each risk factor:

$$c_i = \sum_{j=1}^{n} a_i q_{ij}, \quad i = 1,2,...,m, \quad (1)$$

where $n$ – number of score estimates' ranges; $m$ – number of risk factors.

For estimation convenience, it is necessary to perform normalizing of average risks (1), which will allow to operate them ($\alpha_i$) as probabilities:

$$\alpha_i = \frac{c_i}{\sum_{i=1}^{m} c_i}, \quad i = 1,2,...,m, \quad (2)$$

The second group of estimates reveals opinion of the customer's experts about each $k^{th}$ provider from the providers group $K$, considered as potential service providers. For each provider, experts assign a score according to the discrete scale from 1 to 5 (see column $b_i$ in the Table 1). Then integral risk $r_k$ for the $k^{th}$ provider is determined as follows:

$$r_k = \sum_{i=1}^{m} \alpha_i b_i, \quad k = 1, 2, \ldots, K, \qquad (3)$$

For data provided in the table 1, the integral risk value is $r_k = 0.71$, which can be interpreted as a minor risk, since it can lie within the range from 1 to 5. However, in case the purpose of estimates calculation is the selection of an alternative provider, then the absolute value $r_k$ is of no significance, since the variant with its maximum value is selected. In such case, it is possible to apply both a normalized variant for the analysis ($\beta_i$ as the risk factor relevance coefficient) and the combined effect of weight and relevance of the risk factors as an integral contribution of factor in the risk:

$$\gamma_i = \alpha_i \beta_i, \quad i = 1, 2, \ldots, m, \qquad (4)$$

For the considered set of initial data, values of weight, relevance and contribution for each risk factor are illustrated on the figure 1.

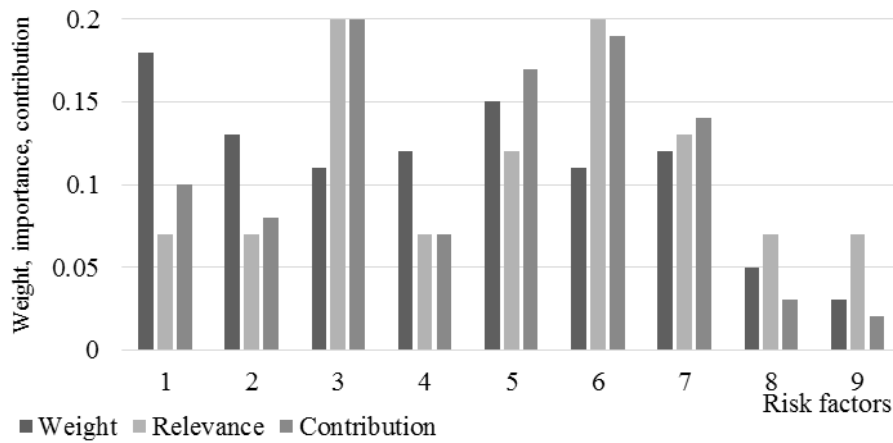

**Fig 1. Values of weight, relevance and contribution of risk factors**

**Conclusions**

The suggested algorithm represents a structured scheme of monitoring of outsourcing providers involvement risks. It can be implemented within the DSS framework. Along with that, expert evaluation of the first group is being updated with involvement of external experts quite rarely, since they demonstrate general situation on the market of such services. Expert evaluations of the second group performed by the customer's specialists, shall be arranged regularly as long as the task of service provider selection is being solved.

Therefore, customer enterprise can ensure high efficiency of the taken management decisions at early stages, using the suggested algorithm of risk assessment of providers involvement.

**References**
[1] D. Brown The Black Book of Outsourcing: How to Manage the Changes, Challenges, and Opportunities / D. Brown, S. Wilson. - John Wiley and Sons, Ltd, 2005. – p. 384.
[2] I. Oshri, J. Kotlarsky, L.P. Willcocks The Handbook of Global Outsourcing and Offshoring. - Palgrave MacMillan. - 2015. - p. 366.
[3] W. Milberg, D. Winkler Outsourcing Economics: Global Value Chains in Capitalist Development. - Cambridge University Press. - 2013. - p. 376.
[4] M.G. Hassana, A. Ojeniyib, M.R. Razallia Practices project management strategies in outsourcing best practices. - Jurnal Teknologi. - January 2016. - DOI: 10.11113/jt.v77.6113
[5] B.Bahli, S.Rivard Validating Measures of Information Technology Outsourcing Risk Factors. - Omega: The International Journal of Management Science. - 2005. - Vol. 33. - pp. 175-187. - DOI: 10.1016/j.omega.2004.04.003
[6] J. Xu, H. Huang The Analysis of Logistics Outsourcing Risk: A System Dynamics Approach. - Logistics Systems and Intelligent Management, 2010 International Conference 9-10 Jan. 2010. - Harbin: IEEE. - Vol.1. -pp.

156-159. - DOI: 10.1109/ICLSIM.2010.5461446.

[7] J.Y. Liu, A.R. Yuliani Differences Between Clients' and Vendors' Perceptions of IT Outsourcing Risks: Project Partnering as the Mitigation Approach. - Project Management Journal. - 29 DEC 2015 DOI: 10.1002/pmj.21559

[8] M. Alexandrova Risk Factors in IT Outsourcing Partnerships: Vendors' Perspective. - Global Business Review. - October 2015 vol. 16 no. 5. - pp. 747-759. - DOI: 10.1177/0972150915591427

[9] B.A. Aubert, M. Patry, S. Rivard, H. Smith IT Outsourcing Risk Management at British Petroleum. - IEEE: System Sciences, 2001. Proceedings of the 34th Annual Hawaii International Conference. - 6 Jan. 2001. - Maui, HI, USA. - DOI: 10.1109/HICSS.2001.927191

[10] F. Ahmed, L.F. Capretz, M.A. Sandhu, A. Raza Analysis of risks faced by information technology offshore outsourcing service providers. - IET Software. - 2014. - Vol. 8. - Iss. 6. - pp. 279-284. - DOI: 10.1049/iet-sen.2013.0204

[11] Z. Luo, J. Wang, W. Chen A Risk-Averse Newsvendor Model with Limited Capacity and Outsourcing under the CVaR Criterion. - Journal of Systems Science and Systems Engineering. - March 2015. - 24(1). - DOI: 10.1007/s11518-015-5263-3

[12] S. Dadelo, Z. Turskis, E.K. Zavadskas, R. Dadeliene Integrated multi-criteria decision making model based on wisdom-of-crowds principle for selection of the group of elite security guards. - Arch Budo. - 2013. - 9(2). - pp. 135-147. - DOI: 10.13140/RG.2.1.4560.3360

[13] D. Nordigarden, J. Rehme, S. Brege, H. Walker Outsourcing decisions - the case of parallel production. - International Journal of Operations & Production Management 34(8). - July 2014. - pp. 974-1002. -DOI: 10.1108/IJOPM-06-2012-0230

[14] S.J. Baxendale Outsourcing opportunities for small businesses: A quantitative analysis. - Business Horizons. - 47/1. - 2004. - pp. 51-58. - DOI: 10.1016/j.bushor.2003.11.008